\documentclass[sigconf]{acmart}
\usepackage{makecell} 
\usepackage{multirow}
\AtBeginDocument{%
  }

\copyrightyear{2025}
\acmYear{2025}
\setcopyright{cc}
\setcctype{by-nc-sa}
\acmConference[CHI '25]{CHI Conference on Human Factors in Computing Systems}{April 26-May 1, 2025}{Yokohama, Japan}
\acmBooktitle{CHI Conference on Human Factors in Computing Systems (CHI '25), April 26-May 1, 2025, Yokohama, Japan}\acmDOI{10.1145/3706598.3714325}
\acmISBN{979-8-4007-1394-1/25/04}

\begin{document}



\title[Sensing Movement]{Sensing Movement: Contemporary Dance Workshops with People who are Blind or have Low Vision and Dance Teachers}

\author{Madhuka De Silva}
\email{madhuka.desilva@monash.edu}
\orcid{0000-0002-3110-0990}
\affiliation{%
  \institution{Monash University}
  \city{Melbourne}
  \country{Australia}
}

\author{Jim Smiley}
\email{jim.smiley@monash.edu}
\orcid{0000-0003-3492-6453}
\affiliation{%
  \institution{Monash University}
  \city{Melbourne}
  \country{Australia}
}

\author{Sarah Goodwin}
\email{sarah.goodwin@monash.edu}
\orcid{0000-0001-8894-8282}
\affiliation{%
  \institution{Monash University}
  \city{Melbourne}
  \country{Australia}
}

\author{Leona Holloway}
\email{leona.holloway@monash.edu}
\orcid{0000-0001-9200-5164}
\affiliation{%
  \institution{Monash University}
  \city{Melbourne}
  \country{Australia}
}

\author{Matthew Butler}
\email{matthew.butler@monash.edu}
\orcid{0000-0002-7950-5495}
\affiliation{%
  \institution{Monash University}
  \city{Melbourne}
  \country{Australia}
}

\renewcommand{\shortauthors}{De Silva et al.}

\begin{abstract}

Dance teachers rely primarily on verbal instructions and visual demonstrations to convey key dance concepts and movement. These techniques, however, have limitations in supporting students who are blind or have low vision (BLV). This work explores the role technology can play in supporting instruction for BLV students, as well as improvisation with their instructor. Through a series of design workshops with dance instructors and BLV students, ideas were generated by physically engaging with probes featuring diverse modalities including tactile objects, a body tracked sound and musical probe, and a body tracked controller with vibrational feedback. Implications for the design of supporting technologies were discovered for four contemporary dance learning goals: learning a phrase; improvising; collaborating through movement; and awareness of body and movement qualities. We discuss the potential of numerous multi-sensory methods and artefacts, and present design considerations for technologies to support meaningful dance instruction and participation.


\end{abstract}



\begin{teaserfigure}
  \includegraphics[width=\textwidth]{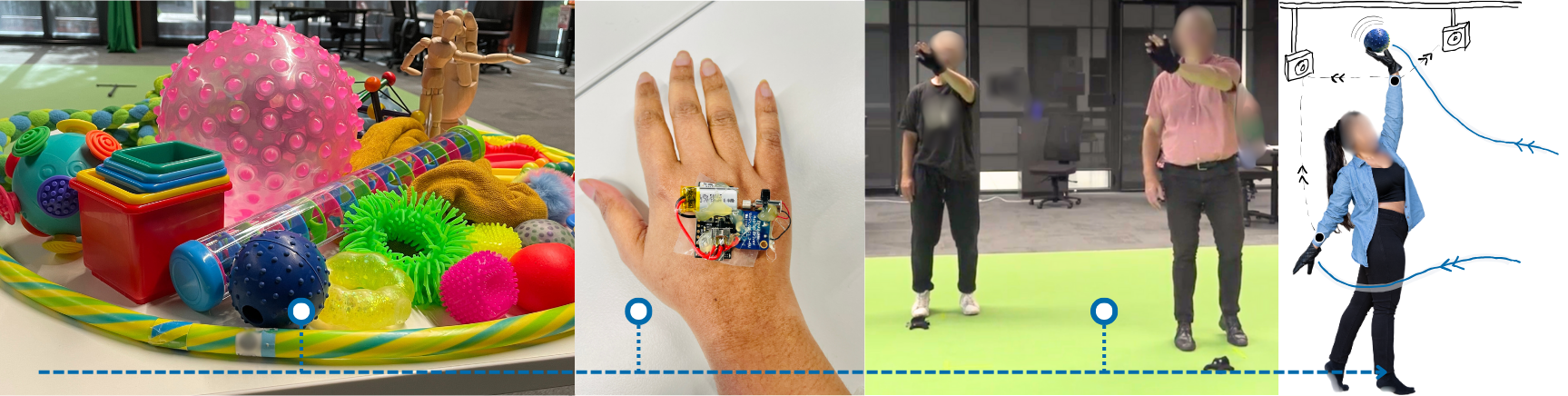}
  \caption{Images from left to right indicate the workshop probes leading to potential future design of accessible dance systems. First image represents the tactile probes, second refers to the haptic feedback probe, third indicates a moment of synchronisation in workshop 3 relying on the sound-based probe, and finally a depiction of future with a dancer being tracked by a motion capture system to gain haptic feedback and play with sound emitting tactile artefacts.}
  \Description{The first image from left represents the tactile probes including boxes in different sizes, squishy balls and hand bands, a large rubber ball with nodules, wooden human full body and hand models in portable hand-held sizes, a multi-sensory rain tube that has a spiral structure inside with small sounding balls. The second image refers to the haptic feedback probe, which is attached on top of a hand by double-sided adhesive tape. The third image indicates a moment of synchronisation in workshop 3 relying on the sound-based probe. The teacher and student of workshop 3 are standing side by side holding their left hand worn with a glove, in the air across their body towards the right side. The final image is a depiction of the future with a dancer being tracked by a motion capture system to gain haptic feedback and play with sound emitting tactile artefacts. The dancer is wearing gloves on their hands with one hand up in the air holding a sounding ball and the other hand lowered slightly away from the body.}
  \label{fig:teaser}
\end{teaserfigure}


\maketitle

\newcommand{\hl}[1]{{\textcolor{blue}{#1}}}
\renewcommand{\hl}[1]{#1}

\section{Introduction}

Accessing aesthetic and recreational body movement programs and activities, such as dance, is challenging for people who are blind or have low vision (BLV). 
In teaching dance, instructors typically guide students by having them replicate demonstrated actions ~\cite{duggar1968}. For BLV people, however, visual cues are largely ineffective. Instead, they rely heavily on verbal instructions from their teachers ~\cite{seham2012}. These verbal cues can often be unclear ~\cite{seham2012, esatbeyoglu2021}, and the instructional approach may create challenges in comprehending the movements ~\cite{bisset2016}. Alongside verbal instructions, educators also use physical interactions 
to teach body movement to BLV students ~\cite{duquette2012, oConnell2006, seham2012}. However, this method can sometimes result in misunderstandings and may not be suitable for students who are sensitive to touch ~\cite{oConnell2006, holloway2022}. 


The accessibility of physical activities such as sports and dance, for people who are BLV, can be enhanced by the use of multi-sensory technologies. For example, technologies are explored for reflection of \textbf{limb movement} in floor-volleyball\footnote{a variation of traditional volleyball played on the floor, often adapted for athletes with disabilities, where players maintain contact with the ground while hitting the ball over a lower net.} learning through a refreshable graphics display~\cite{kobayashi2021}, for the prediction of \textbf{direction} in goalball \footnote{a team sport played by BLV people on the floor with a ball that emits sound, allowing players to track it by hearing.} training using pre-recorded sounds~\cite{miura2018}, and for understanding the \textbf{timing} of a tennis ball in an exergame based on vibro-tactile feedback ~\cite{morellivi-tennis2010}. Although prior research has focused on similar characteristics (movement qualities) of body movement applicable to dance as well, limited research has been conducted exploring how multi-sensory technologies could support dance learning and teaching.


In this research, we focus on Contemporary dance. Contemporary dance provides an excellent style for exploration as it primarily considers movement qualities such as space and effort as the key aspects, following the path of several notable pioneers such as Laban~\cite{laban1971mastery, laban1974effort}. It also considers a collection of many dance genres, cultures and techniques \cite{Jurgens2021}. Finally, Contemporary dance has the ability to connect with other habitual body movement patterns such as everyday walking, exercise, and play that may not overwhelm novice BLV dancers. 
Considering these dance related factors and the identified research gap with limited accessible dance focus, this work aims to explore \textbf{how technology may support Contemporary dance learning through multi-sensory technologies}. This is explored through three main research questions:
\begin{enumerate}
    \item How can different sensory stimuli (tactile, auditory, and haptic) contribute to the dance learning process of BLV people?
    \item How can the use of multi-sensory probes affect or inspire dance teaching techniques? 
    \item How might the insights gained inform design considerations for future multi-sensory systems for supporting engagement with Contemporary dance?
\end{enumerate}

To explore these questions, we conducted five dance workshops, each with a different BLV student and dance teacher pair, with a number of sensory probes inspired by the sensory bodystorming of Vidal et al. \cite{vidal2018}. \hl{These probes—tactile, auditory, and haptic—were designed to inspire movement learning and creativity: the tactile probes are objects with varied textures, weights, and shapes; the auditory probe is motion-capture-based, mapping movement to sound changes; and the haptic probe is a vibro-tactile feedback-based wearable device, with an additional controller enabling one person to cue another through vibrations.} The design and choice of the probes were significantly influenced by the experience and expertise of Author 1 (a dancer with three years' experience with Contemporary dance and improvisational dance workshops) and Author 2 (a former professional musician with over 25 years' experience, as well as expertise in audio technology, motion capture, electronics and fabrication). Through a semi-structured procedure, each student-teacher pair explored the probes through dance instruction, and later through the development of a collaborative, improvised new dance phrase (movement sequences). 

Qualitative analysis of the workshop and post-activity reflective interview revealed five main design considerations 
and \hl{14} sub-design considerations, 
framed around four dance goals that emerged as the main themes from the analysis of the workshops: learning a phrase; improvising; collaborating through movement; and awareness of body and movement qualities. Key novel insights 
include the supportive role of sonification\footnote{Translating movement parameters into sound.} (2B, 2C, 4C) 
for the dancer and teacher to be in sync while learning a phrase or improvising with each other, and 
the importance of multi-sensory artefacts supporting improvisation by integrating existing frameworks and creating dance scores\footnote{A set of guidelines or prompts used to inspire spontaneous movement.} with BLV dancers (4A, 4B, 3C). \hl{This work uniquely contributes to assistive technology and HCI research through insights into the role sensory technologies can play in supporting the accessible teaching and learning of established dance techniques and practices. This research prioritises the ways these technologies can enrich the embodied experience of learning dance.}

\section{Related Work}
While there is research considering recreational body movement \hl{including competitive sports and athletics}, and work examining how technologies have been designed for dance learning or practice, research is limited in the context of exploring dance technologies for BLV people. The following sections introduce dance education and practice for BLV people, then reviews relevant work through the lens of designing solutions in the BLV context, specifically regarding the design of solutions in different non-visual modalities.

\subsection{Body Movement and Dance Educational Practices for BLV People}

The conventional methods for conveying motor skills include pictorials, verbal descriptions, and mirroring the teacher ~\cite{adams1987, bandura1997}. 
The common teaching approach of saying ``do it like this'' 
is the least helpful for BLV people~\cite{duggar1968}. Seham~\cite{seham2012, seham2015, haegele2019} emphasises the need for clearer communication of desired movements and actions when working with this community. However, adjusting appropriate language to accommodate BLV students is a significant commitment of time and expertise from educators~\cite{zitomer2016}, posing a considerable challenge. Our findings from previous work~\cite{desilva2023} and also related literature~\cite{rector2017} suggest the use of metaphors when describing movement, which is a strength for low vision students and blind students who lost their vision in later stages of life but can be challenging for congenitally blind students 
~\cite{desilva2023}.
Physical interaction is another important teaching technique 
for training BLV people in body movement \hl{~\cite{duggar1968, paxton1993, oConnell2006, haegele2019, Larsson2006}}. To establish a verbal vocabulary for movement, it is often necessary to use touch, such as allowing a student to feel the teacher's movement (tactile modeling) or be moved by the teacher (physical guidance)~\cite{duggar1968}. These methods have inherent benefits and challenges: they may not be suitable for students who are sensitive to touch ~\cite{oConnell2006}; can be intrusive to the personal space of the instructor~\cite{esatbeyoglu2021}; \hl{and it can be difficult to demonstrate the complexities of movement ~\cite{desilva2023, albarran2021}. Albarrán et al. has tried to understand the combined role of verbal and tactile communication for learning classical dance, as experienced by both teachers and BLV dancers. They highlight that these communication methods can sometimes fall short in effectively conveying the nuances of dance movements,  
which can lead to potential misunderstandings in the learning process.} 


\subsection{Tactile Tech for Body Movement Learning}

Traditional methods for representing non-visual data include verbal descriptions and raised-line drawings~\cite{BANA2010, Landau2013, RoundTable2022}. However, given the dynamic, three-dimensional nature of body movement, other tools, such as manipulative human models are also used by some teachers to provide tactile feedback ~\cite{claire2023, desilva2023}. 
Simpson and Taliaferro~\cite{simpson2021} discuss the practical implications of using 3D prints to support physical education and \hl{Sucharitakul et al. ~\cite{Sucharitakul2023} have explored 3D printed tactile models to teach the fundamental poses of traditional Thai dance to BLV adolescents.} 
However, there is limited discussion as to how BLV adults perceive the use of 3D objects in compared to BLV children or adolescents ~\cite{desilva2023}. Another tactile approach to representing body movement 
is the use of refreshable graphics displays 
but it has some limitations such as 
problems with showing multiple limb movements ~\cite{holloway2022} and requiring the user to move away from the space of action to refer to the displays
~\cite{desilva2023}.

\hl{Beyond accessibility, movement-based design research ~\cite{Vega-Cebri2023, Tajadura-Jimenez2022, vidal2018, bang2023, Segura2019} has shown that tactile probes can inspire creative body movement, such as Tajadura-Jimenez et al.’s body transformation wearables supporting physical activity through bodystorming ~\cite{Tajadura-Jimenez2022}. However, research exploring tactility for supporting accessible dance learning or teaching is limited. 
Similarly, the role of materiality in this context is also under-explored, although it has been investigated for hand movement in accessible making \cite{Das2020, Zhang2024, Recupero2021}. Das et al. \cite{Das2020} highlight how material qualities such as texture and tension in weaving enable BLV weavers to intuitively understand system states, detect mistakes, and shape designs through tactile feedback. Building on earlier work, it is worth exploring materiality, focusing on the performative character of materials (``what materials and their properties make us do'' \cite{vidal2018}) to support learning or teaching body movement for BLV people.}

\subsection{Auditory Tech for Body Movement Learning}

Commonly used sound-based traditional and low-tech tools in body movement training for BLV people are whistles, clapping, and beeping balls~\cite{haegele2019, desilva2023}. 
Recording sounds for reference has been explored in the past literature and is also being used by some teachers as a technique. For example, some teachers reported in our prior studies that they record movement and ask their BLV students to listen to the sound as feedback ~\cite{desilva2023}. A similar approach was considered in some desktop applications for body movement training \hl{such as works by Miura et al~\cite{miura2018, miura2018b} developing a goalball training application using recorded binaural moving sounds to support the prediction of the direction, height, and distance of an approaching ball based on the sound.} To date, regarding spatial awareness, many technology-driven, auditory-based solutions have focused on the movement of an external object~\cite{miura2018, watanabe2022, miura2018b, sadasue2021, cooper2022, doherty_dont_2022}, and on a person's full body orientation 
such as running \cite{folmer2015, rector2018}, skiing \cite{Miura2023} and swimming \cite{oommen2018, muehlbradt2017}. 
\hl{For example, Goby \cite{muehlbradt2017} serves as an additional tool to offer verbal cues (feedback for corrective action) to swimmers, alerting them when they veer off course 
and Folmer \cite{folmer2015} explains how the sound of drones can be equipped to support blind runners which presents an alternative to verbal cues.} 

Spatial awareness also considers the movement of a person's body parts beyond full body orientation. Work in this context has primarily focused on activities such as yoga~\cite{rector2017}, rock climbing~\cite{ramsay2020}, and dance~\cite{dias2019, Katan2016}. \hl{Dias et al.'s ~\cite{dias2019} work aimed at supporting web-based dance education experimenting with blindfolded users through a sound synthesis technique. In their work, recording-based audio instructions act as directional cues on how to perform specific movements, while real-time feedback provides immediate auditory responses based on the learners' motions, allowing learners to assess and correct their actions in relation to the teacher's recorded demonstrations. The challenge of their work lies in its reliance on initial recordings by the teacher, which limits direct interaction and collaboration in co-located learning environments. Additionally, the system has not been tested with BLV people, despite its potential as a solution for remote learning.} In Katan's study \cite{Katan2016}, sonification is used to translate BLV dancers' movements into sound \hl{in a co-located environment} and revealed problems in conveying nuances of direction, specific limb movements, and poses. While space is an important dimension in dance education, particularly in Contemporary movement \cite{laban1971mastery}, 
it remains unclear what level of detail is necessary in dance movement and how such information can be aesthetically represented for BLV people.

\subsection{Haptic Tech and Multi-sensory Tech Body Movement Learning}

Haptics have been used in several directional ~\cite{aggravi2016} and feedback \cite{junior2022, rector2018} mechanisms in body movement training of BLV people, especially fully-body localisation \cite{junior2022, rector2018}. For example, \hl{Júnior et al. \cite{junior2022} evaluated a solution for BLV athletes participating in Paralympic 100m races, which consists of vibe boards to be placed on the left and right arms, to indicate directional adjustments to stay within lane boundaries (feedback for corrective action). In a similar study, 
Rector et al. \cite{rector2018} 
evaluated the precision and user satisfaction of walking with a human guide, listening to pre-recorded verbal instructions, feeling wrist vibration, and hearing head beat.} 
Some exergames \cite{morellivi-bowling2010, morellivi-tennis2010} have explored how continuous vibro-tactile sensations can aid in understanding the spatial and timing aspects of body movement. 
Several other studies compare or combine the haptic feedback with other modalities such as human voice and sonification. For example, Aggravi et al. ~\cite{aggravi2016} designed a haptic assistive bracelet that allows ski instructors to provide left/right \hl{directional guidance} 
to their students.

\hl{Haptics in dance learning has been explored broadly. For example, Camarillo-Abad et al. used haptic feedback to facilitate leader-follower interactions by conveying directional cues in remote~\cite{CamarilloAbad2018} and co-located dancing~\cite{CamarilloAbad2019}. However, these systems focus on structured dancing, but offer limited support for dynamic cueing, free movement, or inexperienced dancers. Other studies have explored haptics for accessible dance experiences, such as people with hearing impairments \cite{Shibasaki2016} and deafblind audiences \cite{Hossny2015}, emphasising broader interpretations of performance \cite{Shibasaki2016}, often relying on participants with strong performance backgrounds \cite{timmons2019}.}


While there is work that explores how multi-sensory technologies can be used to support different forms of body movement instruction, there is little research exploring it in the dance context, particularly in \hl{free movement exploration, such as Contemporary dance}, 
and in working closely with BLV people \hl{of varying dance experience} and teachers \hl{in co-located environments}.

\section{Method}

Five ideation workshops were organised, each involving one dance teacher and one BLV adult engaging in Contemporary dance while exploring the use of multi-sensory artefacts as part of instruction and improvisation. The workshops were held in a large mixed reality studio room consisting a large square-shaped floor and motion-capture cameras and speakers mounted on the ceiling (Figure \ref{fig:soundprobe}).


\subsection{Participants}

\textbf{Dance Teachers} were recruited through the first author's connections and reaching out through social media. The dance teachers have varied backgrounds, years of experience, and familiarity with teaching BLV people. Five teachers expressed interest and workshop availability. Eleven \textbf{BLV adults} indicated interest in participating in the study, with five selected and matched to a teacher based on availability and level of experience. They represented a diverse range of ages, genders, blindness conditions, and dance experience (Table \ref{tab:demo}). All participants were compensated for their time spent during the workshop, travel costs, and for the teachers' workshop preparation time.

\begin{table*}
\caption{Demographic information of participants. W = Workshop, S = Student, T = Teacher. G: Gender, F: female, M = male, N = non-binary. Y = years of teaching experience. Body Movement Experience refers to students' experience, or teachers' professional experience and dance practice of interest. T4*: has taught dance to five people who have low vision, T5** has a close family member who was blind.}
\label{tab:demo}
\Description{The workshops consisted of a teacher-student pair representing different ages, genders, blindness conditions, body movement experiences and teaching experiences. Workshop 1 consisted of a 45-54 year old legally blind since 17 female student who has done Yoga, Tai chi and dance before. Workshop 1 was guided by a 25-34 year old female teacher with more than 12 years of teaching experience. Workshop 2 consisted of a 25-34 year old congenitally totally blind female student who has done fitness activities before. Workshop 2 was guided by a 25-34 year old non-binary teacher with more than 11 years of teaching experience. Workshop 3 consisted of a 45-54 year old congenitally totally blind male student who has done competitive social ballroom dance before. Workshop 3 was guided by a 45-54 year old male teacher with more than 25 years of teaching experience. Workshop 4 consisted of a 25-34 year old congenitally legally blind and now totally blind female student who has been in theatre before. Workshop 4 was guided by a 45-54 year old female teacher with more than 33 years of teaching experience. She has taught dance to five people who have low vision. Workshop 5 consisted of a 25-34 year old congenitally legally blind non-binary student who has done fitness activities before. Workshop 5 was guided by a 25-34 year old female teacher with more than 10 years of teaching experience. She has a close family member who was blind.}
\begin{tabular}{lllllll}
\toprule
\textbf{W} & \textbf{S/T} & \textbf{Age} & \textbf{G} & \textbf{Vision condition} & \textbf{Body Movement Experience} & \textbf{Y} \\ \hline
W1 & S1 & 45-54 & F & Legally blind since 17 & Yoga, Taichi, Dance &  \\
 & T1 & 25-34 & F & Sighted & Contemporary, Yoga, Pilates & 12+ \\ \hline
W2 & S2 & 25-34 & F & Congenitally totally blind & Fitness activities &  \\
 & T2 & 25-34 & N & Sighted & Contemporary & 11+ \\ \hline
W3 & S3 & 45-54 & M & Congenitally totally blind & Competitive social ballroom dance &  \\
 & T3 & 45-54 & M & Sighted & Contemporary, Contact Improvisation & 25+ \\ \hline
W4 & S4 & 25-34 & F & \begin{tabular}[c]{@{}l@{}}Congenitally legally blind \\ and now totally blind\end{tabular} & Theatre, Ballet &  \\
 & T4* & 45-54 & F & Sighted & Contemporary, Ballet, Theatre & 33+ \\ \hline
W5 & S5 & 25-34 & N & Congenitally legally blind & Fitness activities &  \\
 & T5** & 25-34 & F & Sighted & Contemporary, Acro & 10+ \\ \bottomrule
\end{tabular}

\end{table*}

\subsection{Researcher Positionality}

\hl{While the involvement of BLV people and professional dance instructors was central to the engagement and co-creative process in this work, it is also important to acknowledge how the researchers’ positionality influenced the design of workshop activities and the scope of the study. In particular, authors 1 and 2 played key roles in shaping the probes and activities based on their personal and professional experiences.}

\hl{Author 1 drew on three years of Contemporary dance training and a lifelong exploration of diverse dance styles to inform the probe designs. Experiences in improvisation workshops shaped their ideas, such as incorporating tension bands for physical exploration and using textured balls inspired by an activity imagining a ball moving across the body, blending prior research on tactile stimuli with a focus on novel textures to enhance engagement.}

\hl{Author 2 drew on over 25 years of experience as a musician and audio technologist to shape the sound and haptic elements. Their expertise in audio recording and motion capture guided the creation of sound probes that linked musical elements to movement, with a priority on conveying positional information subtly, musically, and only when necessary. 
Collaborative sessions in which both authors experimented with sound and movement mappings refined the probes, balancing technical precision with creative expression.}

\subsection{Workshop Structure, Pre-engagement and Participants’ Engagement}

\begin{figure*}
  \centering
  \includegraphics[scale=0.29]{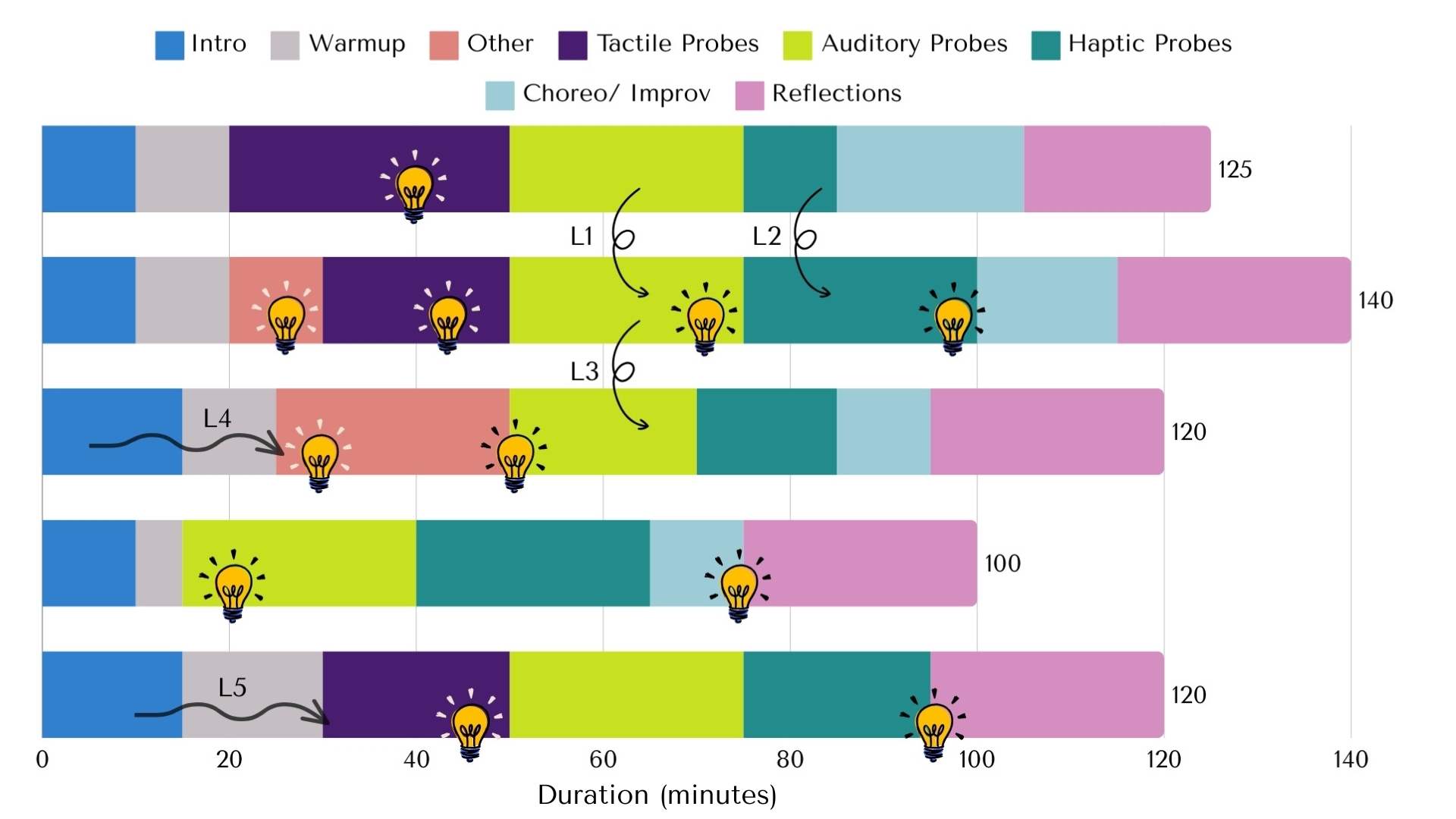}
  \caption{Workshop structure showing time spent on activities during the workshop (legend indicating the activity), idea generation instances (light bulbs), \hl{inter `Links' (L) occurred between the workshops (L1-L3), and intra `Links' occurred within a workshop (L4, L5). Durations are rounded to the nearest 5 minutes and breaks are not shown.}}
  \Description{Workshop structure showing time spent on pre-workshop engagement (diagonal stripes indicating the online engagement) and activities during the workshop (legend indicating the activity), idea generation instances (light bulbs), inter links occurred between (L1-L3) the workshops and and intra links occurred within (L4, L5) a workshop. Interms of pre-workshop engagement, W1 did not have pre-engagement sessions, W2 had an in-person engagement of 2 hours, W3 had an in-person engagement of 1.5 hours, and W4 and W5 had online 30 minutes engagement. All workshops had similar time spent on introductions and warmup, except W4. Workshops 2 and 3 indicate a variation after the warmup compared to other workshops where those teachers explored frameworks and techniques before moving with probes. There were some instances with intense brainstorming of ideas in all workshops. In W1, it was midway during object play. In W2, it was during technique exploration, and towards end of each probe exploration. In W3, it was during framework exploration, then connecting it with the sound probe and midway in the haptic probe exploration. In W4, it was at the start of the sound probe introduction and towards the end of the haptic probe exploration. In W5, it was during the object play and towards the end of the sound and haptics combined probe exploration. Some links occurred between and within the workshops. There were two instance of W1 linked to W2 from sound and haptic probes explorations. Then another from W2 to W3 based on the sound exploration. Interms of links within the workshops, there were links from introductions in W3 and W5 to subsequent sections framework exploration and object play exploration accordingly. W1 and W2 spent time learning a choreography or a phrase after all probe explorations but in other workshops it happened spontaneously. Finally, all workshops had similar reflection duration.}
  \label{fig:brainstorm}
\end{figure*}

Each workshop lasted for between two and three hours. 
The five workshops broadly followed the same structure, as illustrated in Figure \ref{fig:brainstorm}: participant self-introductions, describing their dance or other body-movement related experiences (10-15 minutes); a teacher-facilitated warm up (5-10 minutes); exploration of the artefacts, categorised as tactile objects, sound-based, and haptic sensory probes (15-20 minutes each); learning a phrase or engaging in improvisation (10-15 minutes); final reflections on the workshop and technology (15-20 minutes). 

\hl{The researchers’ involvement in the workshops was limited to presenting the probes, offering minor encouragement and guiding the sessions when needed, allowing BLV participants and instructors to lead their exploration. The dance teachers acted as facilitators, deciding what and how to teach using the probes. The researchers occasionally provided clarifications in response to direct questions from the participants, for example regarding the goal of interacting with the probes, or technical constraints of the system. The research team also provided technical support when adjustments to the probes were needed. For example, when a BLV participant found the initial sound mapping unintuitive, they modified the probe to better align it with the participant’s expectations. 
This collaborative approach ensured the probes remained flexible and responsive to the participants’ needs while maintaining the workshop’s co-creative focus.} 



All teachers had an understanding of the workshop agenda and basic details of their students such as blindness condition, dance experience and specific needs. For example, one participant with low vision preferred contrast with the background; therefore, we accommodated them by asking the teacher and the facilitator to wear lighter shades of clothing to contrast with the black background. There was a pre-engagement session with the teachers before each workshop with the exception of the first workshop. Two of these sessions occurred in person, while the other two took place online, aligned with the teachers' preferences and convenience. During these sessions, they were provided with details of the probes. Two teachers (T2, T3) were able to try the sound-based probe during the pre-engagement session and another teacher (T4) just before the workshop. T5 only had a conceptual understanding of all the probes.


Figure \ref{fig:brainstorm} presents a visual representation of the flow of each workshop, \hl{highlighting how participants 
shared their ideas (see the light bulbs in Figure \ref{fig:brainstorm}). The differing workshop timelines (Figure \ref{fig:brainstorm} and Table \ref{tab:timeline})} show that while each followed the same basic structure, the flow changed between workshops, directed by the interests and exploration of the dance teacher and BLV participants and key insights could emerge at any time. In some workshops (W2, W3) the brainstorming of ideas even occurred before the probe explorations. 
\hl{For instance, in W3, there was frequent play and improvisation guided by T3.} 
The tactile probe (object play) was limited to three workshops (W1, W2, and W5) due to time constraints. 
In some workshops (W2, W3), teachers spent time introducing students to 
improvisation frameworks and scores, later integrating them with the probes (See `Other' in Figure \ref{fig:brainstorm}). 
\hl{There are some intra-links within the same workshop (L3-L5) and inter-links between the workshops (L1 and L2). For example, some teachers were inspired by students' introductions: T3 incorporated an improvisation experience as S3 mentioned that they cannot do improvisation (L4) and T5 recreated a mosh pit experience during tactile probes (object play) based on S5's only experience of dance (L5). The workshop probes were also inspired by participants; S1 suggested expanding the sound probe for further creative movement (L1) and an additional probe (vibration controller, Figure \ref{fig:hapticprobe}) was designed and added to the haptic probes based on an idea by T1 (L2). Furthermore, the auditory probe was expanded with additional speakers based on S2's suggestions (L3)}.

\begin{table*}
\caption{\hl{Duration of workshop participation and pre-engagement. W = Workshop, h = hours, m = minutes, C = Choreography or learning a phrase, I = Improvisation or play. Except for pre-engagement, all durations given in minutes and rounded to the nearest 5. Breaks are not shown.}}
\label{tab:timeline}
\begin{tabular}{@{}llrrrrrrrrl@{}}
\toprule
\textbf{W} & \textbf{Pre- engage} & \multicolumn{1}{l}{\textbf{Intro}} & \multicolumn{1}{l}{\textbf{Warmup}} & \multicolumn{1}{l}{\textbf{Other}} & \multicolumn{1}{l}{\textbf{Tactile}} & \multicolumn{1}{l}{\textbf{Auditory}} & \multicolumn{1}{l}{\textbf{Haptic}} & \multicolumn{1}{l}{\textbf{C/I}} & \multicolumn{1}{l}{\textbf{Reflect}} & \textbf{Total} \\ \midrule
W1                & -                    & 10                                 & 10                                  & 0                                  & 30                                   & 25                                    & 10                                  & 20                               & 20                                   & 125            \\
W2                & 2h (inperson)        & 10                                 & 10                                  & 10                                 & 20                                   & 25                                    & 25                                  & 15                               & 25                                   & 140            \\
W3                & 1.5h (inperson)      & 15                                 & 10                                  & 25                                 & 0                                    & 20                                    & 15                                  & 10                               & 25                                   & 120            \\
W4                & 30m (online)         & 10                                 & 5                                   & 0                                  & 0                                    & 25                                    & 25                                  & 10                               & 25                                   & 100            \\
W5                & 30m (online)         & 15                                 & 15                                  & 0                                  & 20                                   & 25                                    & 20                                  & 0                                & 25                                   & 120            \\ \bottomrule
\end{tabular}
\end{table*}

\subsection{Sensory probes}





The following multi-sensory artefacts were presented in the workshops: 

\begin{figure*}
  \centering
  \includegraphics[scale=0.4]{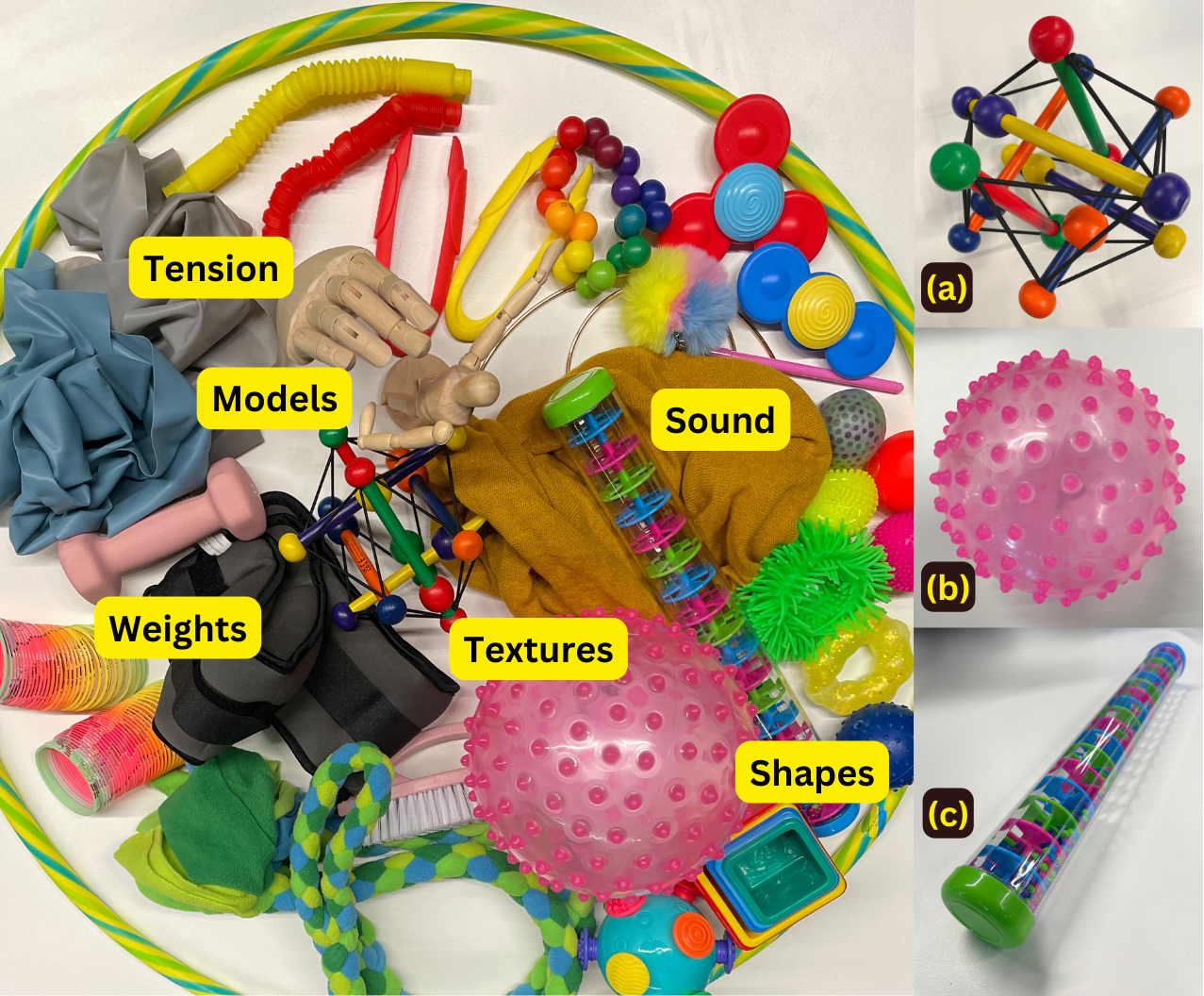} 
  \caption{Objects provided as tactile probes are displayed in the image on the left. Images on the right starting from top, (a) tensegrity structure, (b) a ball with nodules, (c) raintube were frequently used and discussed by the participants.}
  \Description{Four images are provided as a collage. One image provides a top overview of all objects provided as tactile probes. It consists tension bands, a 1kg dumbbell, 1kg foot weights, boxes in different sizes, squishy balls and hand bands, rubber ball with nodules, wooden human full body and hand models in portable hand-held sizes, sensory pop tubes, tongs, a scarf, a cleaning brush, spinners, clothe ropes with handles on both ends, a hula hoop, spiral springs, and interlocking molecule band. Three other images show frequently discussed objects. First is a tensegrity structure with around 16 wooden balls of around 1 inch diameter connected each other through wooden sticks as well as flexible strings, second is rubber ball with nodules, and last is the rain tube, originally designed for multi-sensory play by kids. It has a spiral structure inside where small balls roll through it when the tube is moved while emitting sounds.}
  \label{fig:objectsprobe}
\end{figure*}

\subsubsection{Tactile}
We provided multiple `off-the-shelf' objects (Figure \ref{fig:objectsprobe}) with different material properties such as textures, heavy, tensile, shapes, noisy and friction. This exploration was to spur creativity and to see how tactile sensation can support understanding of movement qualities. This concept and choice of the objects were inspired by self exploration of the first author and prior research. Vega-Cebrián et al. \cite{Vega-Cebri2023} highlighted the importance of using tangible objects in movement-based design methods in facilitating idea generation 
and sensory exploration 
leading to more embodied and experiential outcomes. In particular, considering tactile modality, the objects' affordances can have implications to future design \cite{Vega-Cebri2023}. 
Object selection was inspired by Gong et al.'s \cite{Gong2023} structures (geometric, ability to grab and kinetic), 
Tajadura-Jimenez et al.’s ideation materials (shapes, weights and textures),  Bang et al.'s scarf~\cite{bang2023} (cloth material), 
findings from research team's previous work~\cite{desilva2023} reporting use of sound based tactile tools by teachers (beeping balls), and
self explorations by the first author engaging in improvisation (tensions bands). 

\subsubsection{Auditory}
\begin{figure*}
  \centering
  \includegraphics[scale=0.35]{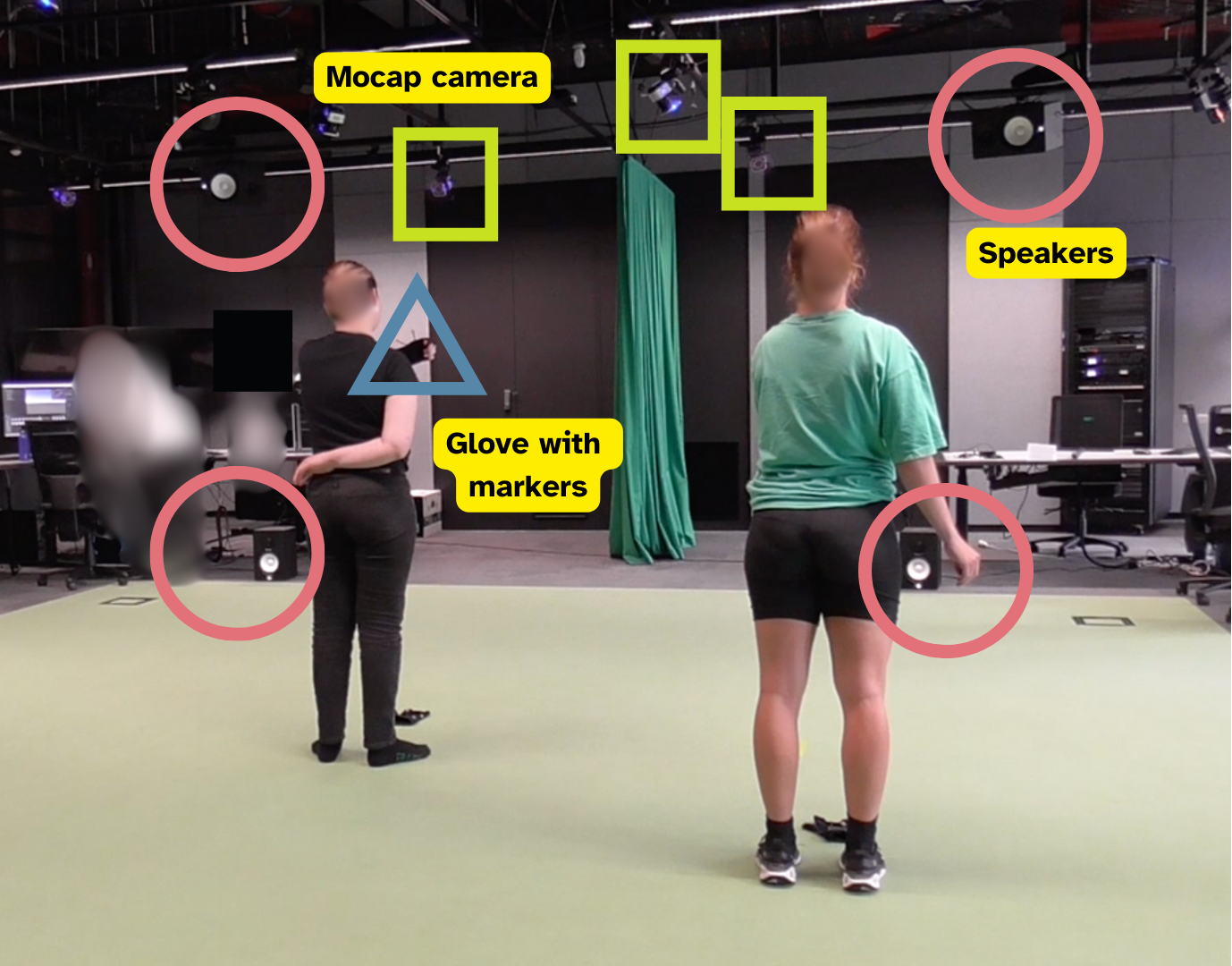}
  \caption{The setup for the sound and haptic probe explorations consisting Motion capture (Mocap) cameras indicated by square frames, speakers indicated by circular frames and tracking gloves worn by student indicated by a triangular frame. The BLV student stands on the left of the teacher who is standing on the right of the image.}
  \Description{The setup for the sound and haptic probe explorations consists of Motion capture (Mocap) cameras and two speakers mounted on overhead rails in a stereo setup and two more speakers on the floor, also in stereo. BLV student stands to the left and slightly in front of the teacher.}
  \label{fig:soundprobe}
\end{figure*}

For this exploration we designed and developed a technology driven probe, mapping elements of a song to body movement tracked by motion capture cameras. This probe was inspired by prior research on movement sonification \cite{Jensenius2012, Tobias2012, Giomi2020}, and exploration by the first and second authors. 
The probe utilised a Vicon motion tracking system and four Yamaha studio monitors: two mounted on overhead rails in a stereo setup and two on the floor, also in stereo (Figure \ref{fig:soundprobe}). Initial tests with spatial audio headphones were conducted, but the spatial audio was found to be imprecise, with significant Bluetooth latency. The streamed Vicon data was processed in an Unity application through a calculation of a vector considering teacher's tracked hands relative to their tracked waist, or in the workshop case, their static position on the floor for simplification. This was then compared with the student's pose, and x and y values for both hands based on the difference in pose were streamed into a MaxMSP application which controlled the music. The stem tracks (isolated instrumental guitar, drum, bass, synthesiser, and vocal recordings) 
of an original song were made available to us by a local sound engineer and their client, a singer, who were also compensated. A further keyboard and organ track was recorded by the author 2 in order to provide additional sonic contrast resulting in a 
guitar track, keyboard track, and a mixed drum,  bass and vocal track. A MaxMSP application synchronised the playback of the three tracks, and using the x and y values, would pan the guitar and keyboard tracks, each attached to a hand, horizontally and vertically. This could be inverted in order to either `pull the sound toward the center' or `follow the sound'.  High and low pass filters were also used for the vertical axis, and binary white noise added on either side of the correct vertical zones. The volume of the main vocal track would fade in and out based on the overall accuracy of the student pose. 
The relatively exploratory nature of the technology was for it to act as a prompt for investigation rather than as a formal evaluation.

\begin{figure*}
  \centering
  \includegraphics[scale=0.18]{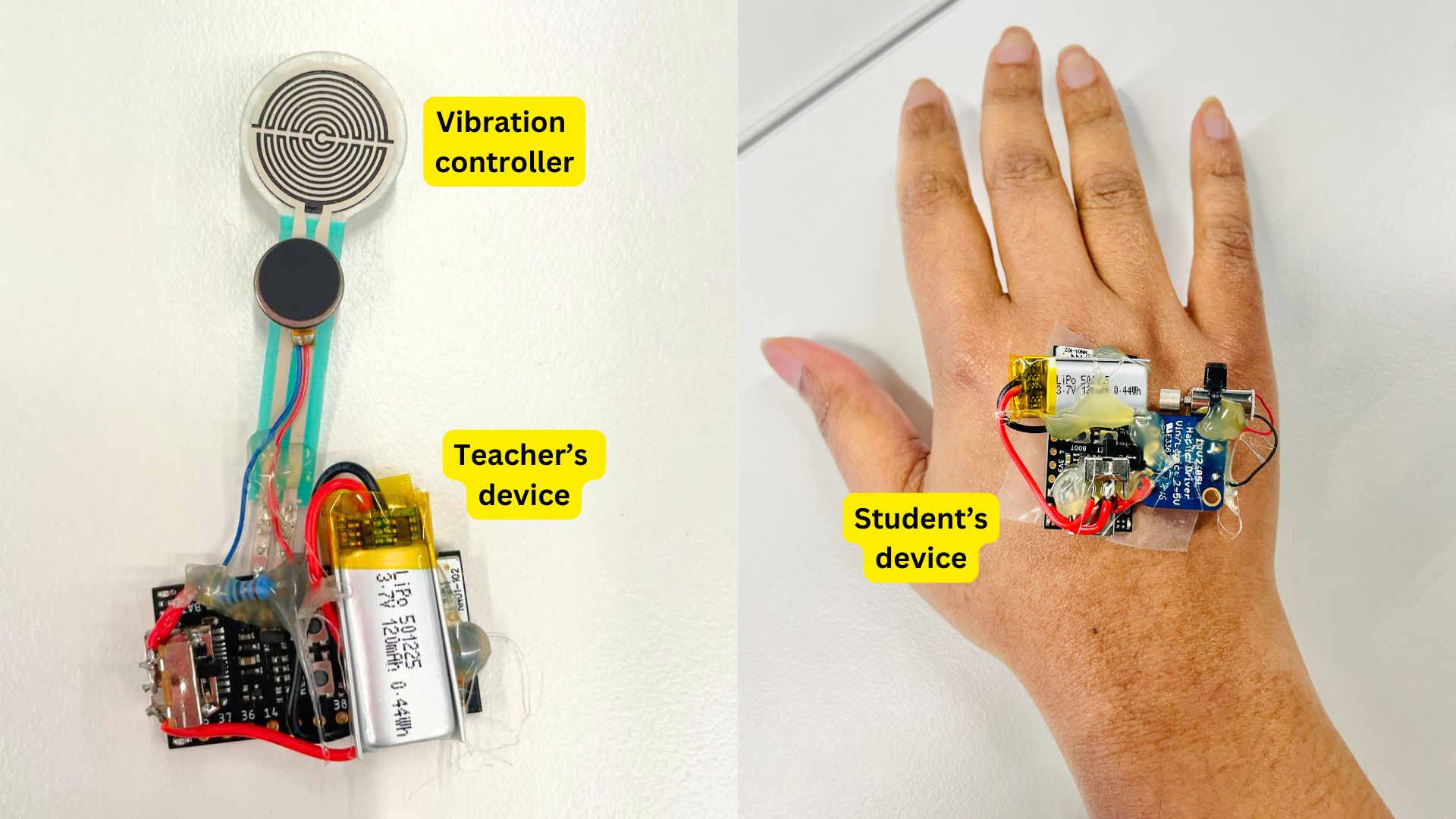}
  \caption{Image on the left represents the vibration controller that allowed the teacher to trigger vibration on the student's device by pressing a circular tactile resistive pad. Image next is the haptic device that lets the student feel both the feedback based on tracked data and teacher's cues based from the controller.}
  \Description{Two images are presented side by side. One image represents the vibration controller with a circular tactile resistive pad to control the vibration including a vibration motor to feel the vibration by the teacher. The other image shows the haptic device that lets the student feel both the feedback based on tracked data and teacher's cues based on the controller.}
  \label{fig:hapticprobe}
\end{figure*}

\subsubsection{Haptic}


The potential of using vibrations for real-time guidance and feedback, along with the choice of the mapping, was considered in the haptic probes. `Tactile dowsing' in Bowling \cite{Morelli2010} 
allows BLV players to sense the direction for throwing the ball when 
the player's hand moves closer to the target direction, through the vibro-tactile feedback being more continuous. This inspired the possibility of exploring continuous vibrations for spatial understanding. The same application used in the auditory probe also controlled the haptic device, via wifi UDP for low latency. The haptic probe consisted of a LIPO battery, esp32 based micro-controller board, and a vibration motor (student's device in Figure \ref{fig:hapticprobe}). The haptic and auditory probes were explored both simultaneously and asynchronously. 
Lu et al. \cite{Lu2023} used vibro-tactile technology 
in a sensorial bodystorming setting for musical instrument learning by BLV people. Inspired by this (and based on Workshop 1 insights), a second probe was added, which allowed the teacher to trigger vibration of the student's device by pressing a resistive pad. 
We explored the possibility of giving the teacher control of the vibrations (teacher's device in Figure \ref{fig:hapticprobe}) instead of preset with vibration patterns \cite{Lu2023}. 

\subsection{Data Analysis}

\subsubsection{Overview}
All workshops were recorded with two static cameras facing each other and recorded audio to document discussions. The transcripts and video recordings were subjected to a thematic analysis based on a qualitative approach, using both inductive and deductive coding in several rounds. \hl{The full thematic analysis including video analysis was conducted in MAXQDA software \cite{MAXQDA_2024}.} A set of initial codes was devised by the research team, with two researchers coding all transcripts and video clips in the first round of coding. During this process, several other codes were identified by both researchers. Three researchers discussed the coding, reconciled differences, and defined the final codebook. One of the initial coders then recoded all transcripts based on this codebook, with the second coder conducting a final validation round on a subset of the transcripts. The research team met to discuss emerging ideas and used these insights to propose the design considerations.

\subsubsection{Inductive and Deductive Coding}

\hl{For analysis of the transcripts and video recordings, overarching high-level codes were first deduced from the research questions (RQ1 and RQ2) considering the probes and teaching techniques
(e.g., `Tactile probe', `Auditory probe', `Haptic probe', `Verbal guidance', and `Physical guidance'). Any further codes were inductively formed during the analysis, some of which were then allocated to the deductive high-level codes. For instance, `sound as patterns', `placing on body' and `vocal effects' were inductively identified but placed under `Auditory probe', `Haptic probe', and `Verbal guidance' respectively. Some of the codes were inductively identified (not linked with the deductive codes) and later were formed into themes, such as `observing movement' and `teacher or student as the leader', which were grouped under the `Supporting Collaboration' theme.} 

\subsubsection{Video Analysis}

\hl{Further to 3.5.2, videos were analysed to enrich the understanding of participants' statements and actions, as well as to clarify instances where the teacher and students had different recollections of their activity. "The videos were coded into segments using the timeline feature of the MAXQDA tool. Video coding allowed the identification of interactions that were more visible in video, such as collaborative activities and artefact manipulations.} 

 \section{Results}
 \label{resultslink}
Here we present the results of the thematic analysis, with the themes primarily relating to the use of the workshop probes. Themes relating to the use of the probes for specific aspects of dance instruction are presented, followed by how they support higher-level aspects of dance, such as collaboration and improvisation. Finally, reflections on how the probes supported the goals of the teacher, enhanced their overall dance experiences, and the implications for the design of technology are presented.

\subsection{Supporting the Development of Body Movement Concepts}

\subsubsection{Body awareness}

Several comments were made on how \hl{the structure of the \textbf{objects} provided as tactile probes} is useful in building body awareness and getting to know the body (T2 and T5).
T2 and T5 shared that the objects were essential in the process of linking movement to the body. For instance, T5 explained, ``Without that object, we wouldn't have produced that movement. Like that process was essential. But it doesn't need to be about the object. It's about the body''. 
T2 also used objects to explain body anatomy, particularly fluid movement, ``It's called a tensegrity object (Figure \ref{fig:objectsprobe}a). You'll notice these beautiful little wires and little wooden pieces. 
Even though none of the wooden bits are touching, the pressure that the wire gets kind of makes the object maintain its shape. 
So the human body is very similar to this". \hl{T2 revisits this object later while explaining how to move in fluidly while bending knees, stating, ``We're not kind of snapping into our joints or our ligaments, but it's a kind of soft strength, like that tensegrity like the beautiful shape if that makes sense''. In response, S2 demonstrated the fluid nature through a knee-bending action while manipulating the tensegrity object through contraction and release, stating, ``So we are doing this [tensegrity object coiled/ contraction] and this [tensegrity object back in shape, released]''. }

\subsubsection{Conveying Tension and Textures}

The affordances of \hl{some of the tactile \textbf{objects}} were frequently referenced by T1 and T2 in explaining force and texture-related concepts. T1 mentioned a key challenge they experienced while using the tension band, ``The hardest thing I find when describing resistance is to get that tension in the body''. Affordances were used in multiple ways, one of which involved exploring the affordance and recalling it, such as T1 instructing participants to feel the tension in a stretch band and then replicate the movement without it. 
Sometimes, T1 moved objects along the student's body, such as a ball with soft nodules (Figure \ref{fig:objectsprobe}b) to allow S1 to feel the textures, ``
If we were to turn this into movement, 
imagine that the sphere is kind of moving across us in, I will roll it across you... 
but then because of these nodules, this is not smooth as the elastic band''. 
To convey force and textures, T2 used \textbf{physical guidance} in several instances, one being 
letting S2 feel a handshake \hl{to further explain joint and muscle movement while exploring the Tensegrity object (Figure \ref{fig:objectsprobe}a), ``I'm going to shake hands now [T2 holds S2's hand]. And I'm going to try to translate to the same quality as we found in the shape, 
I'm going to try to see if you can feel it through my body. 
I'm gonna go through three stages, I'm going to be really sloppy. Then I'm going to call electricals, the toxic masculinity handshake''. T2 then asked S2 to replicate the same movement, which demonstrated S2's understanding of how movement quality varied from soft and relaxed to strong and tense.} 

\subsubsection{Continuity and Timing}

T2 tried to explain continuity using weights \hl{during (\textbf{objects}) exploration by asking S2 to hold a weight (2kg ankle weight) and move while carrying it in their hand} ``
We want to use momentum continuously to keep sequencing it together. And someone's holding something a bit heavier can help us find that continuous sense of momentum 
\hl{[S2- yeah][...] it exaggerates the natural weight of the body''.} 
In learning the timing and duration of movement, S5 found the feedback from the \textbf{sound probe} useful \hl{for better understanding of what `slow' meant when T5 conveyed the speed of the movement. S5 reflected on this thought after T5 tried teaching a phrase that consisted of different speed variations of the movement while experimenting with the sound probe, }``
I was able to hone in more effectively on how fast I should move because I don't know like my `quickly' might be faster than T5's but having that Sonic helps me to try and match 
\hl{that, I think, without that guidance it was quite tricky to know exactly what I was missing}''. 
However, in W4, for understanding timing of the \hl{dance phrase} S4 relied on T4's counts, \hl{for instance, T4 instructing,  ``Are you happy to listen to my tempo? So let's say we're here raise the arm to the left 1, 2, 3, 4, keep it there one there and then go down. 1234... ''}. 
\hl{During the post-reflection, } when S4 was asked about their perspective on the addition of timing elements in the system 
they responded against it, ``that would be overkill, \hl{there's some people out there who say, absolutely, put up as much as you want}''. \hl{In W2, S2 needed time-based cues when learning the dance phrase taught by T2. One of the needs was knowing when to start the next movement, which was supported by T2's cues through the \textbf{vibration controller}. In another instance, S2 required to know how fast a certain movement has to be which was cued by T2} \textbf{vocalising} their hand movement from shoulder to front, making a sound similar to a windy sweep.

\subsection{Probes in Conjunction with Verbal Guidance}

\subsubsection{Specificity}
\label{sec:Specificity}
Some BLV students (S1, S3, S4) shared how they appreciate the specificity when teachers instruct on body movements. They shared the level of specificity they expect, as S3 stated, ``Do you use your hand, or just the finger? It'd be so specific for me, I would need to know''. 
When describing a phrase or choreography, \textbf{linking to the body} in the details, rather than to a spatial location, seemed intuitive. All teachers used this method of detailing a \hypertarget{linktobody}{dance phrase}, with T4 and T5 using it prominently, stating, for example, ``Place our left hand on our head... Once you you find the belly button...From here, slowly come down to left knee...''. \hl{This was observed mostly when teachers were attempting to teach a phrase using the sound probe.}

\subsubsection{With Metaphors}
\label{sec:metaphors}
All teachers used \textbf{metaphors} in their instructions. While teaching a phrase, T1 used several metaphors to describe certain actions such as explaining a push action, ``Right hand presses forward and out. Think this as either opening a gigantic door or like a big field of grass pushing your way through'', and describing shapes, ``Through the middle, the hands going up the wall over the top of the wall and then down the other side, so big waves finger tips reaching up, over down''. The metaphors predominantly provide a sense of tactile imagination, as S1 explained, 
``They are quite visual but they are also tactual, it gives you that bodily sensation that you know what you're going for. Not just `put your leg forward or back' ''. Some metaphors were less relatable to the BLV dancers, such as when T3 used a spatial metaphor during an improvisation session, ``...we had to have things that we used to remember. So here are your top shelf drinks'' which was confusing for S2, as they explained that, for them objects are placed in close proximity.  

\subsubsection{Vocal Effects}
\textbf{Vocal sounds} were used by teachers in different ways, such as adding a sonic effect to their instructions (T1,T2,T3) and as directional and temporal guidance (T2,T3). \hl{Some of the vocal sounds used by T1 to describe movements in the dance phrase resembled the sound of wind blowing while stating, ``I am kind of punching the air'', to wind being drawn in while stating ``Right elbow gets picked up...'', and to some weight being released while stating ``Its like hitting a bubble and rebounce''.} 
T2 also used sound effects \hl{in a similar way but further extended them to convey the speed of the movement they were about to perform while teaching a phrase to S2, vocalising}, ``321.. Whoosh''. T3 introduced S3 to the space by taking a creative stance, using his voice as a beacon to provide direction.

\subsection{Probes in Conjunction with Physical Guidance}

There were several instances in which teachers used physical guidance \textbf{in conjunction with verbal guidance}. \hl{During the warm-up activity, T2 spent some time explaining the fluidity of movement using a \hypertarget{metaphorwave}{metaphor} and while also physically guiding S2}. 
\hl{First, T2 instructed S2 on body positioning and the movement to explore, stating, ``I want you to rock your hips from side to side...as you continue doing this, I want you to get lazier and lazier...'', 
and then T2 rubbed S2's shoulder to hand, trying to replicate a wave-like form on their skin, stating,} ``I want you to find some peacefulness like lighter, like treading water, like moving side to side''.
\hl{Next, T2 spent more time extending physical guidance to explore improvisation with S2 before moving to the object exploration.} For instance, T2 asked S2 to react to their hand placements on S2's body, ``I will place my hand [T2 keeps a palm on S2's shoulder], 
\hl{[...]I'm going to put some pressure there [T2 pushes S2's right hand back and touches S2's left hand and back simultaneously], moving different body parts}... 
you got the choice now, you move towards the touch or away from it. There's no right or wrong''. 

There were some discussions on \textbf{preference for physical contact} in dance. T2 and S3 suggested that the probes could be useful for people who are reluctant to engage in physical touch. T2 said, ``if you are mapping different parts, it might be really useful. And especially likely to better represent people that don't want to be touched''. 
T2, T3, and T4 (\hl{the teachers who used physical contact as part of their guidance}) sought consent whenever they physically contacted their students. \hl{T2 engaged in physical contact more frequently than the others and established this agreement at the outset, ``I'll use my hand [T2 holding S2's hand] I'll ask you before I touch each time, I'll let you know when it's happening, or if you are not comfortable at any point, let me know''.}

\subsection{Supporting Improvisation and Play}

\subsubsection{Teaching/ Learning to improvise}
\label{sec:improvi}
During T3's attempts to introduce S3 to improvisation, S3 asked for details of the movement a couple of times 
but T3 either avoided providing details in some instances or suggested alternative movements. In addition, T3 guided S3 in trying unconventional moves, ``
Now let's swap it, [S3- oh!] let's go out with the arms, but in with the feet. So reverse it. So we're just challenging our habits''. T3 took this initiative based on an earlier conversation with S3, in which S3 stated, ``I could never improvise, I can copy you. But I cannot create as much as I love art...'' \hl{(L4 in Figure \ref{fig:brainstorm})}. 
After this activity, T3 incorporated movement qualities into the improvisation. 
\hl{T3 prompted S3 to create different shapes with their hands and then T3 moved their body responding to S3 (S3 acting as leader similar to a music conductor) producing sounds such as loud thuds by jumping or vocalising, ``You could like wrinkle your hands or wave your hands around... You could go as your expression of movement. And I'll try to do it with my voice lit up''. After a while, T3 modified the prompt, asking S3 to now improvise by responding to sounds made by T3, ``Can we reverse it? So I'm the leader. So, to my sound, you respond through your body movement''.}

Moving with or without \hl{the tactile} \textbf{objects} for conceptual play was explored by T1, who emphasised the role of tactile engagement beyond verbal imagination, ``
to have tactile things, so I am not just describing `imagine'''.
T5 used multiple objects to create sound (moving the rain tube, scrubbing the brush on the floor), probing S5 to respond through movement while repeatedly saying, ``you can do traditional or reject them''. For instance, S5 responded to those cues by rolling on the floor with the rubber ball (Figure \ref{fig:objectsprobe}b). 

\subsubsection{Dance Frameworks and Scores}

All teachers brainstormed how scores\footnote{A dance score is a set of instructions or notations that describe how a dance should be performed. Scores can be highly detailed or more open to interpretation, depending on the choreographer's intent.} could be created using all probes through different approaches. Starting with \hl{tactile} \textbf{object play}, T1 and T2 were observed recalling the affordances of some objects. T2 stated, ``When it's soft, I want you to find really the tensegrity object (Figure \ref{fig:objectsprobe}a) earlier, find the softness? Can you find that in your body''. Through these explorations, T2 suggested working with particular objects over the long term to better understand their potential qualities, ``
Today, we've come into these objects for the first time, 
let's say we just worked with the tensegrity for a year. I guess the potential for understanding keeps developing around it.''

The \textbf{sound probe} helped the teachers ideate beyond its existing features. T2 suggested exploring an aural soundscape, ``
If you can use this technology to shift someone's relationship to an aural soundscape that gives a blind person for example, something to play with an element in the space that they can then make creative choices to whether that was going to be like performance or class''. 
T3 integrated Trisha's Locus framework\footnote{Trisha Brown's "Locus" is a dance framework that involves an invisible cubic structure around the dancer, where specific points within the cube are numbered. Each letter of the alphabet corresponds to a number, allowing dancers to "spell" words using their movements\cite{locusapplication2020}.} with the sound probe. T3 first introduced the \hypertarget{trishaimprov}{framework} using metaphors, then mapped the spatial locations of Locus to those of the sound probe, and spelled out their and S3's names to explore the framework. The sound probe was useful for S3 in understanding that they had hit the correct locations.

Focusing on the \textbf{haptic controller}, T2 felt more creative with multiple vibration points and suggested that they could create a score for it,  ``Like 
an improvisation score. If there was multiple points, I can put like a framework''. Apart from envisioning a haptic controller with extensions, teachers (T2, T3) used it as-is, providing randomly timed cues and exploring the intensity of the haptic pulse as suggested by S2, ``Maybe what if you press like this one [S2 imitated the press like a pulse]?''. Addressing this idea, T2 then pressed in pulses and suggested a score, ``When you don't feel anything we have to be in stillness. When it's soft, I want you to find the softness''. 

\subsection{Supporting Collaboration}
\label{sec:collab}

With the \hl{tactile probes} (\textbf{objects}) play, collaborative movement was observed most frequently in W5. For instance, S5 suggested moving together with the hula hoop, and, following that idea, T5 created a rule for trying the hula hoop together, ``At all points somebody has to be inside the hoop, but only one''. Later, T5 modified the rule, saying that both could be inside the hula hoop, ``There's an opportunity for two people to be in at once that's okay''. 
They tried a similar movement together, holding the two ropes from opposite sides. 

\textbf{Sound and haptic feedback} were used by BLV students as a means to understand their teacher's movement. 
Observing teacher movements using the probes was visible in W2-W4. S2 first reacted, stating, ``It was beautiful to feel that I was able to kind of move the way T2 moved''
and then later reflected, ``
Sometimes when we're in the right spot, you were moving your arm. 
Oh, no, you moved. So I need to look for it again. 
I kind of had an idea when you were moving your body''. 
As a way of supporting co-creation, S3 used the probe feedback to determine whether their teacher (T3) hit the spatial location they commanded relating to Trisha's Locus dance framework (explained in \hyperlink{trishaimprov}{\textit{Improvisation}}). 
The teacher (T3) expressed satisfaction in being able to listen to the student's feedback, 
``I was very satisfied that we hit it and 
it's good to know direction 
instead of just this sort of vagueness''. 
T3 continued referring to the same framework by asking S3 to suggest a letter or the destination of movement while T3 set the timing for that movement by triggering the vibration of the \textbf{haptic controller}. 
S2 suggested that they can think of collaborating with another blind person to create a dance using sound and haptic probes, ``I think it will be able to even like, create a choreography with another blind person, without needing someone sighted to kind of like indicate things''. 

\subsection{Supporting the Wider Dance Context}

\subsubsection{Education and teaching choreography}
\label{sec:choreo}
S3 shared that education is important, 
highlighting how it builds confidence in BLV people to enter the dance space, ``The education perspective is really important, because many people I knew, would never have the confidence to just go out and go dancing, 
you might feel too self conscious to do it until you get taught and this is where the systems come in''. 
They further shared use of \textbf{haptic feedback} for posing precisely, ``I will never know when you say lower or higher. But this thing tells me on my hand,
It is a very precision, an empowering thing''. From a teaching perspective, T1 suggested an approach to conduct classes using all probes, ``
I'd be making a class using all of these things, then you could maximise different effects,
part of the class would be all about the sound, and this part will be all vibration''.
Some modifications (T1, T2) were suggested for how the \textbf{sound probe} could be further useful, such as using surround sound, ``If I had access to a room with full surround sound, I could create choreography and movement in that direction'', and  
mapping musical patterns to movement, ``You could use patterns that could become recognisable so they learn and follow along [S1 agreeing]... I can shake my hands and anyone can do that at different speeds''. 


\subsubsection{Supporting Groups} 
While outside the scope of the workshop, some speculated (T1, T2, S2, S4, T5, S5), stating that the sound and haptic probes might be helpful in a group setting. T1 was keen on it based on the premise that they might need more support if they had to teach someone who is totally blind in a group setting, ``
We are still kind of using vision here whereas if you had blind people, this could have been much more complex, in a class setting, if I have someone with no vision, and to do what I do in a normal class, it would be almost impossible''. From a student's perspective, S2 explained how the \textbf{haptic feedback} could help them stay in sync with others, ``
with the vibration, I know when people are changing because otherwise, I'm kind of lost''. 
For similar reasons, S4 appreciated the \textbf{sound probe}, ``I must say, having lost the vision that I had, that told me whether I was in sync with other people, having the white noise thing really helped''. 

\subsection{Participant Preferences}
\label{sec:Preferences}

Preferences were discussed throughout all probe explorations. Among the workshops (W1, W2, W5) that included a session of \textbf{object play}, object selection was determined by either the teacher-student pair together (W1, W5) or solely by the teacher, T2 picked the objects in advance and suugested additional objects prior to the workshop (Tensegrity, Figure \ref{fig:objectsprobe}a). Reasons for preferences included predictable nature (S1), 
 visual appeal preferred by teachers (T1), 
the potential of conceptually mapping to movement, and additional auditory stimuli. For instance, the rain tube (Figure \ref{fig:objectsprobe}c) was brought up in discussion in all three workshops without being prompted by researchers. S1 did not find the rain tube inspiring, but S2 preferred it, stating they like noisy objects, and S3 moved to the sound of the rain tube rolling on the floor by T5.


Preferences for the tech probes were mostly about the mappings (more details in the \hyperlink{map}{Mappings} section). For instance, some reflected that the horizontal mapping was confusing (S1, S4), while others preferred one mapping over the other (S2, S5). 
In terms of having both \textbf{haptic and sound feedback}, S3 reflected that they preferred having both, ``It's more engaging to do both, but also, why create something that only fits one type of learner'', they revealed their recent hearing issues and also commented, stating ``Even if you can hear properly, you might not always focus''. However, S4 expressed a contrasting opinion, ``
I found it extremely distracting. 
It's almost like you're expecting people to listen to too many cues, telling the same thing''. 

\subsection{Technology Implications}

\subsubsection{Teaching with machines}
\label{sec:machines}
\hl{Some teachers (T3, T5) questioned why it is necessary for a system to provide the directional or temporal cues for the BLV dancer when they can instruct them verbally. Some BLV dancers countered this argument, explaining that they do not always receive honest feedback.} 
S3 explained the predictability of a machine compared to a human when providing feedback, ``If it's artificial, like the music, the way we did with a glove that is very predictable, a human is never predictable because they can make a mistake'',
to which T3 responded, ``So an impartial witness, and machinery has impartiality''. From a co-creation perspective, T5 explained how the system can support in creating a 
a sequence that is partly human-created but partly machine-driven, ``I like the idea of trusting the system, 
like we were finding it through improvisation, 
and that could be the choreography to an extent. 
The system could have developed the choreography, so then you escape the system, it's not a part of it anymore''. 

\subsubsection{Mappings}


There were varied comments on the \hypertarget{map}{mapping} that the \textbf{sound probe} offered. 
In W1, the sound probe had mapping only for horizontal guidance (panning), and 
within those limitations, S1 commented, ``I feel the task was quite complex and it resulted in a very basic movement [showing horizontal hand movement]''. 
S4 and S5 also found the panning confusing, ``I'm just having trouble like, I know I'm trying to get things centered, 
it's messing my head, whether I want to move in the direction that the sound is coming from or move to the opposite direction to this''.
From a teacher's perspective, T3 shared a similar opinion, stating the confusion when there is no direct correlation in the mappings,  ``If it was the direct thing, I put my arm up and it'll be be like [vocal sound effect]...what I'm doing is not something I have to read to interpret along’’.
When we switched the mappings (to `following the sound’), S2 and S5 responded slightly faster than with the initial mapping (`pulling the sound’). 

\textbf{Haptic} feedback was appreciated (S2, S3, S5) mainly for its specificity, as S2 stated, ``So then the vibration told me, no, you're not correct, you need to go a little bit to either side or lower or higher’’. Although S5 appreciated the vibration feedback, they found it difficult to perceive at low intensity, ``I find it a bit hard to perceive when it's just slightly rumbling. Am I just feeling like my heartbeat or something?''. In W2, we explored a counter mapping (where vibration occurs closer to the correct position) to which S2 found less intuitive, ``I think this time it was a bit confusing. I thought it was correct because the sound was okay… but then it was still vibrating. Maybe it's not correct. I didn't know what to do''.



\subsubsection{Training needs} 
Many reflected that the probes were new to them and that they would prefer to keep exploring.  S2 shared, ``Maybe if I had experience with these, 
what different sounds mean? Because sometimes you explained me, but I kept forgetting, what it means? ''. S3 shared a similar reflection, ``I struggled with getting the guitar centred. All I could really do was to chase the vocal. But it took a while to, to learn how to do it now, not forever''. 
Some teachers (T1, T2, T4, T5) wished for more time with the sound probe. T4 commented, ``Maybe we have to spend separate time getting used to the system? Like, because we're just exploring that first layer? I definitely feel invested in that''. 


\section{Discussion}

Here, we discuss the five Design Considerations (DC1-5) and 14 sub-Design Considerations (A-D, e.g., DC1A), the main contributions of the study, which help us imagine possible multi-sensory representations either for improving current body movement learning systems or for future considerations when designing tools for inclusive dance teaching and learning practices. The DC are aligned with the four primary dance learning goals: learning a phrase, improvising, collaborating, and building awareness of body and movement qualities, which help form the structure of the discussion (see Table \ref{tab:dc}). Each DC and sub-DC is presented, linking it with dance goals, modalities, and the related themes from qualitative analysis (section \ref{resultslink}), inspired by the presentation of findings in Trajkova et al.'s work \cite{Trajkova2024}. Note that some DCs span multiple dance learning goals (DC3, DC4, and DC5). 

Table \ref{tab:dc} presents an overview of how the goals, modalities, and themes link with the DCs and Sub-DCs. Additionally, it provides links to the most relevant (non-exhaustive) prior work with some annotations to indicate novel insights. We conclude with an overview of reflections on the workshop process.


\begin{table*}
\caption{Overview of the dance goals, modalities, themes and design considerations (DC / Sub-DC) with inspirations from prior work. A11y = Accessibility. * indicates novel sub-DC that are under explored either in accessibility or dance focused research.}
\label{tab:dc}
\setlength{\tabcolsep}{2.8pt} 
\begin{tabular}{|l|l|l|l|l|l|l|l|l|l|l|l|l|}
\hline
\textbf{Dance Goal}                                                                                       & \textbf{Modality}                                                            & \rotatebox{90}{\textbf{Movement Concepts (4.1) }} & \rotatebox{90}{\textbf{Verbal Probes (4.2)}} & \rotatebox{90}{\textbf{Physical Probes (4.3)}} & \rotatebox{90}{\textbf{Supporting Improv. (4.4)}} & \rotatebox{90}{\textbf{Supporting Collab. (4.5)}} & \rotatebox{90}{\textbf{Dance Context (4.6)}} & \rotatebox{90}{\textbf{Participant Prefs. (4.7)}} & \rotatebox{90}{\textbf{Tech. Implications (4.8)}} & \textbf{DC} & \textbf{Sub-DC} & \textbf{Related Work} \\ \toprule

\multirow{3}{*}{\textbf{\begin{tabular}[c]{@{}l@{}}Learning \\ a phrase (5.1) \end{tabular}}}                      & Verbal guidance                                                              & x                                                                    & x                                                         &                                                             &                                         &                                                  &                                                   &                                        &                                        & 1           & 1A, 1B          & Dance: \protect\cite{fisher2022unpeeling, Jones2023}                                                                                                                                     \\ \cline{2-13}

\textbf{}                                                                                  & Haptics                                                                      &                                                                     &                                                           &                                                             &                                         &                                                  &                                                   & x                                      & x                                      & 2           & 2A              & A11y: \protect\cite{Alizadeh2014, cosgun2014evaluation, morellivi-bowling2010}                                                                                                   \\ \cline{2-13}

\textbf{}                                                                                  & Sonification                                                                  & x                                                                    &                                                           &                                                             &                                         & x                                                 & x                                                 & x                                      & x                                      & 2           & 2B*, 2C*      & A11y: \protect\cite{Miura2023, Hoque2023, rector2018, Ross2000, aggravi2016}                                                                                          \\ \hline

\multirow{2}{*}{\textbf{\begin{tabular}[c]{@{}l@{}}Improvisation \\ and Play (5.2) \end{tabular}}}                 & \begin{tabular}[c]{@{}l@{}}Physical and \\ Tactile interactions\end{tabular} &                                                                     &                                                           & x                                                           & x                                        &                                                 &                                                   &                                        &                                        & 3           & 3A              & \begin{tabular}[c]{@{}l@{}}A11y: \protect\cite{paxton1993, gay2020}, \\ Other: \protect\cite{sten2022}\end{tabular}                              \\ \cline{2-13}

\textbf{}                                                                                  & Sonification                                                                  &                                                                     &                                                           &                                                             & x                                        & x                                                &                                                   & x                                       &                                       & 4           & 4A*, 4B*          & Dance: \protect\cite{oppici2020, bang2023, liu2021}                                                                                                                                      \\ \hline

\multirow{2}{*}{\textbf{\begin{tabular}[c]{@{}l@{}}Collaboration \\ (5.3)\end{tabular}}}                                                                     & \begin{tabular}[c]{@{}l@{}}Tactile and \\ Haptics\end{tabular}               &                                                                     &                                                           &                                                             & x                                       & x                                                &                                                   &                                        &                                       & 3           & 3B, 3C*          & \begin{tabular}[c]{@{}l@{}}Dance: \protect\cite{kronsted2021improv}, \\ Other: \protect\cite{Popp2023, Yao2011, Yao2011roperev, Beelen2013}\end{tabular} \\ \cline{2-13}

\textbf{}                                                                                  & Sonification                                                                  &                                                                     &                                                           &                                                             & x                                       & x                                                 &                                                   &                                        &                                        & 4           & 4C*              &\begin{tabular}[c]{@{}l@{}}A11y: \protect\cite{Katan2016}, \\ Dance: \protect\cite{Brown2015}\end{tabular}                                                                                                                                                                                                                          \\ \hline

\multirow{2}{*}{\textbf{\begin{tabular}[c]{@{}l@{}}Movement\\ qualities (5.4) \end{tabular}}} & \begin{tabular}[c]{@{}l@{}}Physical and \\ Tactile interactions\end{tabular} & x                                                                   &                                                           & x                                                           &                                         &                                                  & x                                                 &                                        &                                        & 5           & 5A, 5B          & \begin{tabular}[c]{@{}l@{}}A11y: \protect\cite{turmo2020}, \\ Dance: \protect\cite{kronsted2021improv, jeffries2016Thesis}\end{tabular}         \\ \cline{2-13}

\textbf{}                                                                                  & Sonification                                                                  & x                                                                   & x                                                         &                                                             &                                         &                                                  &                                                   &                                        &                                        & 4           & 4D              & Dance: \protect\cite{francoise2014}                                                                                                                                                      \\ \bottomrule 
\end{tabular}
\end{table*}


\subsection{Learning a Phrase/ Choreography}

\subsubsection{Verbal Guidance}

\begin{itemize}
    \item \textit{DC1: Design \textbf{language prompts} to support teachers in instruction-based teaching, providing a range of options understandable to BLV people.}
    \item \textit{DC1A: Consider exploring \textbf{metaphors} suitable for representing particular actions and movement qualities in Contemporary dance.} 
    \item \textit{DC1B: Consider references to \textbf{body positions} as part of the phrase and language prompts.}
    \item \textit{Themes: Verbal Probes (4.2), Movement Concepts (4.1)}
\end{itemize}

Verbal instructions play a crucial role, serving both to introduce movements and to provide feedback for refining posture, spatial orientation, and movement quality, including speed, texture, tension, and continuity. When guiding BLV students, descriptions that incorporate \textbf{metaphors} and references to the body are particularly effective, offering a nuanced way to convey movement concepts. Fisher \cite{fisher2022unpeeling} introduces the Concentric Circles Model (CCM), a tool for analysing dance movements using metaphors starting with a core unit, which serves as the focal point for exploring instinctive interpretations and associations \hl{(`push action' example in \ref{sec:metaphors})}. 
However, metaphors also need further consideration when it incorporating them into dance instruction with BLV people, as our prior work findings \cite{desilva2023} have highlighted that BLV people may interpret some visually-based metaphors differently. 

In some instances, metaphors required further assistance through objects and physical interactions, requiring a multi-sensory approach in corrective instructions \cite{turmo2018}. \hl{Another key aspect that supports teachers in describing the dance phrases, particularly when engaging with the sound probe, is} \textbf{referencing body positions} (\hl{see examples in \ref{sec:Specificity}}). 
Jones et al. \cite{Jones2023} highlight challenges in interactions between performance artists and prompts generated by GPT-3 \footnote{GPT-3 is an AI language model developed by OpenAI that can generate human-like text based on a given prompt.}, noting a lack of consideration for performers’ physical limits, such as unrealistic instructions that force performers to adapt to the algorithm’s commands. \hl{Insights from our study could inform the development of systems that currently overlook human embodiment, emphasising the importance of improving descriptive methods to enhance inclusion and accessibility}.

\subsubsection{Haptics}

\begin{itemize}
    \item \textit{DC2: Design \textbf{continuous multi-sensory} cues for precise guidance of the movement to be in sync with the teacher or others.} 
    \item \textit{DC2A: Consider providing a choice of \textbf{haptic mappings}, such as varying vibration to consistent or varying vibration to pausing the vibration.} 
    \item \textit{Themes: Participant Preferences (4.7), Technology Implications (4.8)}
\end{itemize}

In learning a phrase, haptics are useful for providing precise feedback. Prior work has explored directional cues in dance \cite{Drobny2010} and similar body movement contexts \cite{Alizadeh2014, cosgun2014evaluation}. However, there are limitations in how vibrations can be mapped and recognised \cite{BACHYRITA2003, Alizadeh2014}. Within these constraints, we experimented with two mappings in our design, inspired by prior work \cite{Alizadeh2014, cosgun2014evaluation, morellivi-bowling2010}: one where vibrations became continuous and consistent upon reaching the desired location \cite{morellivi-bowling2010}, and another where vibrations changed and reduced as participants approached the target position. 
Our findings reveal that BLV participants find the latter approach more intuitive and meaningful. \hl{This underscores the need for customising haptic feedback to individual preferences and exploring the dynamic nuances of vibration intensities through further evaluations with BLV people (or other types of learners) interested in utilising haptic feedback.}


Haptic feedback is preferred as part of multi-sensory feedback rather than in isolation. It is particularly useful for older adults \cite{Alizadeh2014} or individuals with hearing difficulties. \hl{However, this should remain a choice rather than as a default setting, as redundant concurrent feedback could overwhelm the students}. \hl{Another critical consideration is the extent to which haptic feedback supports the teacher, as teachers may inadvertently modify their movements without being aware of the student’s progress. To address this, multi-sensory, tech-based \cite{Drobny2010} systems in dance learning could be enhanced to enable teachers to monitor student progress effectively, a key requirement in physical training domains \cite{turmo2018}}. 

\subsubsection{Sonification}


\begin{itemize}
    \item \textit{DC2B: Consider exploring a choice of distinct \textbf{sound}-to-movement \textbf{mappings} such as synthetic, musical and natural soundscapes supporting auditory perception.}   
    \item \textit{Themes: Participant Preferences (4.7), Technology Implications (4.8), Dance Context (4.6)}
\end{itemize}



Mapping between sound and movement has been a focus of scholarly experimentation for some time. \hl{In our exploration, vertical mapping using white noise was effective and intuitive for spatial direction feedback, while horizontal panning posed challenges due to its subtlety and the need for attentive listening } \cite{BACHYRITA2003}. \hl{This highlights the importance of exploring sound mappings that are both intuitive and distinct, including options such as synthetic and natural ambient sounds \cite{Hoque2023} or distinct auditory patterns (refer to \ref{sec:choreo}).} Additionally, auditory training may enhance the ability to perceive and interpret these sounds, suggesting it as a valuable consideration when developing systems that rely on complex auditory cues \cite{lahav2022perception}. \hl{Further investigation is needed to examine how the interplay between intuitive design and auditory training can enhance sound-based feedback systems designed for spatial orientation and movement learning, particularly in contexts prioritising accessibility.} 

\begin{itemize}
    
    \item \textit{DC2C: Compare the level of \textbf{directional} and \textbf{temporal} guidance required from a teacher and a machine.}

     
    \item \textit{Themes: Technology Implications (4.8), Dance Context (4.6), Movement Concepts (4.1), Supporting Collaboration (4.5)}
\end{itemize}

The necessity of a system providing directional or temporal cues for BLV dancers is sometimes debated, with some suggesting that verbal instructions should suffice. However, BLV students point out that verbal feedback may not always be honest or sufficient (see \ref{sec:machines}). While directional guidance is recognised as particularly beneficial in group learning contexts, receiving individual attention can sometimes be challenging, as shared by BLV people in our prior work  \cite{desilva2023}. Several studies have compared verbal guidance with sound-based and other types of feedback, yielding mixed findings, some prefer sighted guide assistance \cite{rector2018}, others favour sound or alternative feedback methods \cite{Ross2000, Miura2023}, while some advocate for a combination of both \cite{aggravi2016}. \hl{While perfect spatial accuracy is not essential in contemporary dance, the need for impartial and accurate feedback suggests that further exploration of these feedback modalities could provide valuable insights. Developing systems that integrate diverse feedback approaches could better address these needs, particularly in inclusive and group-oriented learning environments.}

\subsection{Improvisation and Play}






\subsubsection{Physical and Tactile interactions}


\begin{itemize}
    \item \textit{DC3: Design \textbf{tactile} and \textbf{haptic} cueing artefacts to support improvisation by BLV dancers.}
    \item \textit{DC3A: Consider exploring artefacts to support \textbf{contact improvisation} with BLV dancers.}
    \item \textit{Themes: Physical Guidance (4.3),  Supporting Improvisation (4.4)}
\end{itemize}

Physical interactions, such as those used in Contact Improvisation (CI) \footnote{Contact Improvisation (CI) is a dynamic movement practice focused on spontaneous, non-verbal communication through touch, weight, and movement, encouraging dancers to explore natural motion while staying connected to their partners and surroundings.} \cite{jackson2022contact}, were incorporated at various stages of the workshops. Physical contact was employed as a guiding method, with participants responding to touch cues by either moving closer or maintaining distance. This approach aligns with similar wearable-based interactions, such as the tag game \cite{sten2022} and ``keep your distance'' \cite{gay2020}, which explored intuitive and comfortable areas for touch input. While physical contact was generally well-received, it was pre-announced each time to signal respect \cite{matherly2014navigating, jackson2022contact, desilva2023}, though this occasionally impacted the spontaneity of movement. Steve Paxton’s exploration of CI with BLV people \cite{paxton1993} demonstrated its potential to enhance sensory awareness and create deep connections through touch, transforming movement into a shared, tactile language beyond visual constraints. Further research could explore how CI can be facilitated with fewer disruptions to spontaneity, maintaining its sensory and communicative richness.

\subsubsection{Sonification}

\begin{itemize}
    \item \textit{DC4: Design \textbf{sonification} artefacts to support improvisation by BLV dancers.} 
    \item \textit{DC4A: Consider exploring sonification artefacts for creating \textbf{improvisation scores} by/with BLV dancers and integrating existing dance frameworks.}
    \item \textit{DC4B: Consider exploring \textbf{tangible} and \textbf{intangible} sonification artefacts when designing for improvisation goals.} 
    \item \textit{Themes: Supporting Improvisation (4.4), Participant Preferences (4.7), Supporting Collaboration (4.5)}
\end{itemize}

An essential aspect of creative movement lies in actively exploring diverse movement possibilities. Oppici et al. \cite{oppici2020} highlight how movement sonification can facilitate this process by prompting dancers to discover novel movements through specific musical elements such as frequency and tempo, which correspond to their motions. The sound feedback appeared to inspire improvisation frameworks like the Locus dance framework\footnote{Trisha Brown's ``Locus'' is a dance framework that involves an invisible cubic structure around the dancer, where specific points within the cube are numbered. Each letter of the alphabet corresponds to a number, allowing dancers to ``spell'' words using their movements. For example, to spell a name, dancers reach corresponding points in the cube with any part of their body. This structure offers both rigor and freedom, enabling dancers to create personalized and inventive choreography within a defined spatial framework \cite{locusapplication2020}\label{note1}}, \hl{offering a balance between providing precise directions and encouraging playful movement exploration. This highlights an area worth exploring further, particularly in how sonification can support diverse and inclusive improvisational practices within creative movement frameworks.}

Another way to use sonification for improvisation is through tangible artefact forms, as some of our participants preferred to play with sound-emitting objects. 
Crafting digital musical instruments could be a potential area for further exploration especially encouraging BLV people to move beyond structured movements, aligning with defamiliarisation \cite{Carlson2019}. \hl{For instance, Trolland et al. \cite{Trolland2024} explored how a dancer with cerebral palsy, utilised a digital musical instrument to transform their movements into sound, creating a dynamic interplay between dance and auditory expression that enriched the performance experience while embracing their unique movement capabilities. Beyond accessibility, this idea} aligns with work explored by Bang et al.\cite{bang2023} and Liu et al. \cite{liu2021}. For example, Liu et al. \cite{liu2021} designed a motion-sensing sonic hoop 
that triggers different sounds inspiring dancers to improvise. 

\subsection{Collaboration}

\subsubsection{Tactile and Haptics}

\begin{itemize}
    \item \textit{DC3B: Explore haptic cueing artefacts for creating \textbf{improvisation scores} by/with BLV dancers, considering multiple cues and different placements on the body.}
    \item \textit{Themes: Supporting Improvisation (4.4), Supporting Collaboration (4.5)}
\end{itemize}

The use of a haptic controller probe opens up possibilities for further exploration of its role in encouraging improvisation and collaboration within creative movement practices. For instance, the integration of leader-follower dynamics \cite{CamarilloAbad2019} and frameworks like the Locus dance framework\textsuperscript{\ref{note1}} inspire how haptic feedback could support fluid role transitions and shared decision-making in movement. \hl{Suggestions to expand haptic cues to multiple points on the body and explore feedback for different body areas highlight opportunities to diversify sensory input and create more nuanced movement experiences.} Such ideas resonate with existing work, such as the acro suit by Popp et al. \cite{Popp2023}, which uses vibrotactile cues to facilitate co-creative practices by enabling partners to interpret and respond intuitively. \hl{Their work and ours demonstrate the potential to enhance improvisation and movement flexibility, particularly in contexts where visual feedback is limited. These insights suggest that haptic cueing could be further explored in conjunction with the creation of improvisation scores that are accessible to BLV people or for use in existing movement practices that rely more on proprioception than vision.}

\begin{itemize}
    \item \textit{DC3C: Design tangible artefacts for co-creation with BLV people using \textbf{motion sensing and force-feedback}.} 
    \item \textit{Themes: Supporting Collaboration (4.5)}
\end{itemize}

\hl{The use of tactile artefacts presents possibilities for enhancing collaboration and creative movement. For example, objects such as hula hoops and ropes (see \ref{sec:collab}) have been observed to facilitate physical connections, encouraging participants to move together and engage in dynamic interactions.} This aligns with the concept of social affordances \cite{de2007participatory, kronsted2021improv}, where physical prompts support interaction and collaboration \cite{chionidou2023Thesis}, and reflects prior studies such as Yao et al.'s \cite{Yao2011roperev} exploration of motion-sensing ropes for collaborative play. Exploring how tactile artefacts, combined with features such as force feedback, can support dynamic and expressive movement practices is an interesting direction for future work. 

\subsubsection{Sonification}

\begin{itemize}
    \item \textit{DC4C: Design sonification artefacts to support BLV dancers in being a \textbf{leader or follower} in co-creative practices.} 
    \item \textit{Themes: Supporting Improvisation (4.4), Supporting Collaboration (4.5)}
\end{itemize}

The workshops revealed two distinct uses of sonification: providing self-reflective feedback and conveying information about a collaborator’s movement (see \ref{sec:improvi}). This aligns with prior work, such as Brown et al.'s \cite{Brown2015} Tango-based interactive sonification system, which integrates individual movements and collective choreography to influence real-time music generation. Collaborative applications, as discussed by Katan \cite{Katan2016}, have explored the potential for BLV dancers to rely on sonification to interact with one another. However, challenges remain in conveying detailed aspects of movement, such as direction, specific limb actions, and poses solely through auditory feedback, often requiring additional verbal or physical intervention. \hl{This raises questions about the extent to which sonification can independently mediate detailed interactions. It could potentially function more effectively as part of a broader system, incorporating a social contract where collaborators agree on interpreting feedback, supported by complementary modalities. Further exploration is needed to refine its role in collaborative dance, balancing auditory strengths with additional cues to ensure greater accessibility and nuanced interaction.}

\subsection{Body Awareness and Movement Qualities}


\subsubsection{Physical and Tactile Interactions}

\begin{itemize}
    \item \textit{DC5: Design with \textbf{tactile} affordances of artefacts to translate movement qualities.} 
    \item \textit{DC5A: Consider exploring \textbf{moving with and without} objects and being physically guided by the instructor.} 
    \item \textit{DC5B: Consider long-term explorations of the \textbf{manipulability} nature of artefacts.} 
    \item \textit{Themes: Physical Guidance (4.3), Movement Concepts (4.1), Dance Context (4.6)}
\end{itemize}

The exploration of conveying textures in body movement revealed several approaches, including physical contact to draw patterns on the body (e.g., waves), tactile modeling \cite{duggar1968}, and using the textures and affordances of objects. Objects were employed creatively to enhance body awareness and movement exploration: moving with the object, replicating the movement after releasing it, or placing the object directly on the body \cite{turmo2020}. As noted by Vidal et al. \cite{turmo2020}, placing objects in unconventional locations can encourage learners to discover new movements they might not have considered. Additionally, the affordances of objects were used to translate movement qualities such as force, fluidity, and continuity, framing dance as a sensory exploration rather than solely self-expression. This perspective encourages learners to engage with their surroundings in ways that empower creativity and movement exploration \cite{kronsted2021improv}. Long-term interaction with objects is also essential, resonating with Jeffries’s work \cite{jeffries2016Thesis} where the evolving use of aging materials, like cardboard boxes, shifted from manipulability to deeper physical engagement. \hl{These insights suggest the potential for further exploration of how objects can support embodied, sensory-driven learning in dance.}

\subsubsection{Sonification}

\begin{itemize}
    \item \textit{DC4D: Explore integrating \textbf{vocalisation} by dance teachers to convey effort qualities of Contemporary dance to BLV dancers.} 
    \item \textit{Themes: Verbal Probes (4.2),  Movement Concepts (4.1)}
\end{itemize}

\hl{Using vocal effects to represent movement qualities offers a dynamic way to enrich the exploration of body movement.} This approach aligns with prior work on sonification through vocalising \cite{francoise2014}, highlighting its potential to add nuance and depth to Contemporary dance learning, particularly for BLV dancers. \hl{Further exploration could investigate how vocalisation might enhance the representation of movement qualities and support more accessible learning experiences.}

\subsection{Workshop Reflections}

A notable aspect of this research is the dynamic nature of the workshop structure. A more fluid approach was designed for students and teachers to explore the prompts through authentic creative practice together, rather than to allow a highly prescriptive workshop structure. It was believed that greater insights could be gained into the role of technology in supporting Contemporary dance instruction and improvisation with BLV students by designing a workshop structure that was itself open to creative interpretation. As such, it is important to reflect on this workshop design and its implications. 

\subsubsection{Prior Engagement}
The first two workshops led us to do things differently regarding pre-workshop engagement. Prior to the workshop, T1 was made aware only of the workshop agenda; however, T2 joined in person before the workshop day and engaged with the probes. This prior engagement allowed them to connect with the probes more deeply and participate in a more prepared ``shared participation'' \footnote{Shared Participation allows participants to suggest methods and respond to group dynamics, providing comfort but risking dominance by one individual.} process, rather than joining more as a participant in a ``fully controlled facilitation'' \footnote{Fully Controlled Facilitation involves the facilitator directing the process, allowing participants to focus on tasks, which creates a safe environment but may limit flexibility and spontaneity.} style \cite{Reidsma2022}. 
As such, pre-workshop engagement was encouraged for the following three workshops. Based on teacher preferences, two teachers joined online and two joined in person for the prior engagement. 
The teachers who joined in person improvised and choreographed movement with the probes more than the others. For instance, T2 taught at least two phrases and improvisation, while T3 explored dance frameworks and improvisation with their students.

\subsubsection{Long-term Engagement}

As T2 pointed out, we may gain further insights if there are long-term workshops 
to explore the same probes. Long-term engagement has shown nuanced results in the dance-based research in general \cite{Felice2021} and body movement research on accessibility with BLV people ~\cite{rector2017}. 



\subsubsection{Workshop Limitations}

Some limitations were placed on the participants. Considering the risks of body movement and the needs of some BLV people, we requested teachers to avoid certain movements such as lying on the floor, rapid turns, and high jumps. However, as S2 suggested, we later allowed this to be an open conversation between teacher and student, which led to a greater variety of movements explored in later workshops. 
Our BLV participants preferred listening surround sound speakers over wearing headphones, stating that they did not want to miss out on environmental information. However, in a prior study, the BLV participants have shared otherwise \cite{rector2015}. 
We had planned to organise some workshops with multiple BLV dancers, and with or without sighted dancers but due to schedule conflicts, workshops took the form of teacher-student pairs. 


The choice of prompts also imposed some limitations. During the workshops, we used only body tracking. We may have gained more insights if we had incorporated additional artefacts as Alaoui et al. \cite{Alaoui2017} suggest, combining multi-modal data to represent effort.  
Furthermore, the technology used for the probes is lab-based and not yet mainstream in dance spaces except in experiments and some showcases. However, 
less mainstream tech could help BLV people be more physically active~\cite{rector2015}. 

\subsection{Limitations}

\hl{The workshops provided a rich set of data, with our analysis focused on understanding the role of multi-sensory technologies in supporting instruction and learning within existing Contemporary dance practices. Given the somewhat narrow focus of this analysis, it should be acknowledged that these data could equally be used to examine other themes in accessible dance, such as gaining new insights into blind ways of interacting during dance. However, such topics, can be the focus of future work. This work is also limited by workshops consisting of only two people: the teacher and the dancer. As such, our findings are limited by the lack of a group-focused learning context. Finally, given the exploratory nature of this work, which aims to investigate the possibilities of assistive technologies, there is an acknowledged lack of formal evaluation of the technology-based probes.}

\section{Conclusion}
Dance access remains an under-explored area in HCI, especially within embodied domains like dance education. In this study, we conducted five dance workshops using sensory probes with BLV adults and teachers to explore and understand the nuances of dance access through non-visual means. Our findings resulted in five primary design considerations that can inform the development of future systems aimed at enhancing dance accessibility. As part of our future work, we plan to further develop and refine the ideas generated during these workshops. Additionally, we see potential to expand this work to include children and other physical activities, broadening the scope and impact of our research. By understanding the core elements of human movement and interaction in these contexts, we can contribute to the design of systems that enhance embodied experiences across diverse populations and activities.
  






\begin{acks}
Our sincere thanks to all the dancers and teachers who participated in these workshops. This work is generously supported by funding from the Faculty of Information Technology, Monash University. Thank you to the sound engineer, Myles Mumford and the singer, Matt Glass for allowing us incorporate their musical work.
\end{acks}



\bibliographystyle{ACM-Reference-Format}
\bibliography{sample-base}

\appendix


\end{document}